\RequirePackage{fix-cm}
\documentclass[smallextended]{svjour3}  

\usepackage{CJKutf8}
\smartqed  
\usepackage{graphicx}
\usepackage{caption}
\usepackage{subcaption}
\usepackage{float}
\usepackage{multirow}
\usepackage{hyperref} 
\usepackage{siunitx}
\usepackage{amssymb}
\usepackage[toc,page]{appendix}
\usepackage{xr}
\usepackage[table,xcdraw,dvipsnames]{xcolor}
\usepackage{booktabs}
\usepackage{afterpage}

\makeatletter
\newcommand*{\addFileDependency}[1]{
  \typeout{(#1)}
  \@addtofilelist{#1}
  \IfFileExists{#1}{}{\typeout{No file #1.}}
}
\makeatother

\newcommand*{\myexternaldocument}[1]{%
    \externaldocument{#1}%
    \addFileDependency{#1.tex}%
    \addFileDependency{#1.aux}%
}
\myexternaldocument{Appendix}
\newcommand{\RNum}[1]{\uppercase\expandafter{\romannumeral #1\relax}}

\begin{document}
\raggedbottom
\title{Characterizing the roles of bots during the COVID-19 infodemic on Twitter 
}
\subtitle{\\}


\author{{Wentao Xu$^1$}
        \and {Kazutoshi Sasahara$^2$}
}


\institute{
\email{myrainbowandsky@gmail.com} \\ \\
$^1$ Graduate School of Informatics, Nagoya University, Japan \\
$^2$ School of Environment and Society, Tokyo Institute of Technology, Japan
}

\date{Received: date / Accepted: date}

\maketitle

\begin{abstract}
An infodemic is an emerging phenomenon caused by an overabundance of information online. This proliferation of information makes it difficult for the public to distinguish trustworthy news and credible information from untrustworthy sites and non-credible sources. The perils of an infodemic debuted with the outbreak of the COVID-19 pandemic and bots (i.e., automated accounts controlled by a set of algorithms) that are suspected of spreading the infodemic.
Although previous research has revealed that bots played a central role in spreading misinformation during major political events, how bots behaved during the infodemic is unclear. In this paper, we examined the roles of bots in the case of the COVID-19 infodemic and the diffusion of non-credible information such as ``5G'' and ``Bill Gates'' conspiracy theories and content related to ``Trump'' and ``WHO'' by analyzing retweet networks and retweeted items. 
We show the segregated topology of their retweet networks, which indicates that right-wing self-media accounts and conspiracy theorists may lead to this opinion cleavage, while malicious bots might favor amplification of the diffusion of non-credible information. 
Although the basic influence of information diffusion could be larger in human users than bots, the effects of bots are non-negligible under an infodemic situation.

\end{abstract}

\section{Introduction}
\label{sec:intro}
``We're fighting an infodemic. Fake news spreads faster and more easily than this virus, and is just as dangerous...'' said the World Health Organization (WHO) Director-General on February 15, 2020 \cite{WHO20200915}. 
Prior to this comment, a large amount of misinformation about the new coronavirus emerged on popular social networking sites (SNSs) and these began to play a major role in the diffusion of misinformation. According to \cite{Brennen}, in terms of information sources, top-down misinformation from politicians, celebrities, and other prominent public figures accounted for 69\% of total social media engagement. Additionally, approximately 60\% of the COVID-19 related information was reconfigured, twisted, recontextualized and reworked on Twitter, while 38\% of the misinformation was completely fabricated. The fact-checking organization, Politifact also pointed out that trustful and mostly trustful news about the coronavirus comprised only up to 10\% of the total information \cite{Politifact} being disseminated. SNS users tend to connect to like-minded users, which is known as the ``birds-of-a-feather phenomenon'' or homophily \cite{Menczer2004,Redner1998}. 
Meanwhile, users also tend to follow influencers and celebrities on SNS who present themselves as information hubs. 
Thus, when an influencer posts misinformation, his or her followers, often with no doubts about the content, tend to believe it and share the post in a homophilic social network.

In addition to human users, previous studies have shown that bots (i.e., automated accounts controlled by a set of algorithms) play a key role in propagating misinformation. \cite{MessiasJohnnatanLucasSchmidtRicardoAugustoRabeloOliveira} discussed how bots engaged with humans to increase their influence. A well-known case in politics is the 2016 US presidential election, during which bots were used to widely diffuse misinformation \cite{AlessandroBessi,Varol2017}. 
It was estimated that of all tweeted links to popular websites, 66\% were shared by bots \cite{PewResearchCenter}. \cite{Ferrara_2020} analyzed 43.3 million English tweets and found the preferences of bots and humans in tweet contents by analyzing n-grams and hashtags. Furthermore, it was suggested that bots may play a critical role in driving the viral spread of content from low-credibility sources and may be able to amplify misinformation \cite{Shao2018}. 
\cite{doi:10.1080/10584609.2018.1526238} found that social bots increased from 7.1\% to 9.9\% when examining German Twitter followers' prevalence and activities of seven German parties before the 2017 electoral campaigns. 
Bots were also used for spreading and amplifying vaccine misinformation~\cite{Broniatowski2018a}.
Research about the COVID-19 infodemic has discovered the information flow pathways between humans and bots \cite{Gallotti2020}. A study of social media manipulation in the 2020 US presidential election characterized the differences between right and left-leaning bot behavior \cite{Ferrara_Chang_Chen_Muric_Patel_2020}. In addition, bot activity is believed to be linked to hate toxicity in denser and more isolated local communities in both the US and the Philippines \cite{Uyheng2020}.

There are several methods for bot characterization and identification. Botometer is a well-known tool for automatically detecting bots based on supervised machine learning models \cite{DBLP:journals/corr/DavisVFFM16,Sayyadiharikandeh2020DetectionON}. Botometer examines six classes of features including profile, friends, social network, temporal activity patterns, language, and sentiment which is further categorized into approximately 1,200 features for a Twitter account. This tool computes a ``bot score'' for each user that ranges from 0.0 to 1.0. The higher the score, the higher the probability that the user is a bot.  
Botometer is a state-of-the-art tool for identifying bots, and a series of studies have used this tool to quantify their online behaviors of bots \cite{Shao2018,Vosoughi1146}. We applied Botometer in our study therefore, to discriminate between bots and humans.

Given this context, an important research question is how bots behaved in the spread of misinformation during the COVID-19 infodemic. To study this, we focused on retweets on Twitter. Retweet is an information sharing behavior by which any user can share messages immediately with his or her followers. A retweet can be both a productive communicative tool and a selfish act of attention seekers \cite{Boyd2010TweetTR}. 
\cite{Cha2010MeasuringUI} found that an popular tweet has either an interesting context or it is produced (retweeted) by an influencer. 
\cite{10.1007/978-3-642-23808-6_2} pointed out that a user’s high popularity does not necessarily imply a high level of influence and vice-versa, indicating that an influencer's popularity and influence are weakly correlated. However, \cite{Hong2011,Suh2010,10.1145/1772690.1772751} considered that a user's contextual information (e.g., social network topology, tweet content, URLs) affects the retweet behavior. In this paper, we used retweets to address information-sharing behavior, shedding light on how COVID-19 misinformation is shared in an information ecosystem where bots live.

Misinformation is classified into several types and conspiracy theory is one type \cite{Douglas2019}. 
The negative effect of a conspiracy theory is to elicit emotions including avoidance, fear, anger, aggression and further result in irrational behaviors \cite{Douglas2019}. As an example, the 5G conspiracy theory was reported on January 22 by a local Belgian newspaper, reporting that a local doctor claimed that 5G might be linked to the coronavirus \cite{TEMPERTON}. In the UK, 5G cell-phone masts came under arson attacks due to this conspiracy theory \cite{Hamilton}. Another version of this conspiracy theory claims that 5G alters people's immune systems and changes DNA structures, thus increasing people's susceptibility to contracting the coronavirus \cite{DisinfoLab,Ketchell}. 
Another popular conspiracy theory targeted Bill Gates, co-founder of Microsoft Corporation, and claimed that Gates supported implanting tracking chips in people under the pretext of a mandatory coronavirus vaccination \cite{JackGoodman2,Jr.}. US political groups were reported as showing a  significant partisan bias regarding this conspiracy \cite{yahoo20200910}; compared to the left wing, the right wing was more inclined to believe in this conspiracy.

To investigate the spreading of misinformation during the COVID-19 infodemic, we focused on the above-mentioned conspiracy theories related to 5G and Bill Gates as  mentioned above. 
For comparison, we also focused on other topics, such as ``WHO'' and ``Trump'' (the 45th US president). 
These keywords were selected because health and political misinformation flourished during the COVID-19 infodemic. Moreover, recent research revealed that Trump was the largest driver of COVID-19 misinformation \cite{Evanega2020}.

In this paper, we first characterize the credible and non-credible bots around the four topics in the retweet networks. We then compare the retweet activities as well as other features in the four topics considered. Our results may help us understand how bots played a role during the COVID-19 infodemic and provide insights into a mitigation strategy.

\section{Data and Methods}
\subsection{Data collection and preprocessing}
\CJK{UTF8}{gbsn}
We used Twitter as a data source to characterize the COVID-19 infodemic. 
We collected 279,538,960 English tweets from February 20 to May 31 by querying COVID-19 related keywords: ``corona virus''; ``coronavirus"; ``covid19''; ``2019-nCoV''; ``SARS-CoV-2''; and ``wuhanpneumonia'' using Twitter's standard search API. We focused on four topics in our analyses: ``WHO''; ``Trump''; ``Bill Gates''; and ``5G'' as noted.
Extracting tweets regarding these topics resulted in a total of 37,219,979 tweets, of which 23,1515,441 (82.8\%) were retweets. The breakdown of this dataset is shown in Table~\ref{table:1}.
Some users in our dataset could be considered malicious and were suspended based on Twitter's spam policy during the period between the date we collected the tweets and the date we computed the corresponding bot scores; These users were, thus, not included in our analyses.

According to a list of non-credible websites released on MisinfoMe~\footnote{\url{https://misinfo.me}} and a list of non-credible news website domains released in \cite{Zimdars2016}, 893 responsive websites from the total of 1,143 domains were collected and used as the non-credible domain list. We also examined a list of trustworthy rated media released by \cite{NELA-GT-2018} and retrieved 30 (all responsive) credible media domains. In addition, we added major science journals \emph{Nature}~\footnote{\url{https://www.nature.com}} and \emph{Science}~\footnote{\url{https://www.sciencemag.org/}} as credible domains. Thus, we obtained a total of 32 credible domains and used these as the credible domain list. Based on the credible and non-credible domain lists, each tweet was labelled as ``credible'' if the tweet included a URL in the credible domain list, and as ``non-credible'' if the tweet included a URL in the non-credible domain list. Then, given a topic, each user was labelled as ``credible'' if the user retweeted credible tweets exclusively and ``non-credible'' if the user retweeted non-credible tweets exclusively. 
In other words, non-credible users are those who posted with URLs from the non-credible domain list at least once but never posted with URLs from the credible domain list. Credible users were similarly defined. 
Note that a user's label can change from topic to topic. For instance, a user is labeled ``credible'' in the WHO topic if the user retweets credible domains, exclusively in that topic, even if the user retweets non-credible domains in other topics.

Furthermore, we used the Botometer API V3 to compute bot scores of users. 
In this study, we used the CAP (complete automation probability) value instead of the raw bot score.
Given a screen name of a Twitter account, the Botometer returns a CAP value ranging from 0.0 to 1.0. 
After multiple trials of analyses, We set 0.54 as the threshold in our analyses. 
This means that a user was considered to be a bot if the CAP value was larger than or equal to 0.54; otherwise, the user was considered to be human. 

Using the above discriminative conditions for credible/non-credible and bot/human, we classified users into five types for analyses: credible humans (CH), non-credible humans (NH), credible bots (CB), non-credible bots (NB), and the other.

\begin{table}[t]
\centering
\caption{Overview of COVID-19 tweets by topic}
\begin{tabular}{llllll}
\toprule
 &
  \begin{tabular}[c]{@{}l@{}}Unique \\ users (U)\end{tabular} & 
  {\color[HTML]{3F3F3F} \begin{tabular}[c]{@{}l@{}}Unique Users\\  with \\ Bot score (US)\end{tabular}} &
  {\color[HTML]{3F3F3F} \begin{tabular}[c]{@{}l@{}}Percentage \\  (US/S)\end{tabular}} &
  {\color[HTML]{3F3F3F} \begin{tabular}[c]{@{}l@{}}\# Tweets \end{tabular}} &
  {\color[HTML]{3F3F3F} \# Retweets} \\ \hline
WHO        & 88,719    & 73,704  & 83.1 & 128,016   & 46,650    \\
Trump      & 1,125,251 & 947,694 & 84.2 & 5,631,459 & 2,322,036 \\
5G         & 67,523    & 55,315  & 81.9 & 97,638    & 31,814    \\
Bill Gates & 94,584    & 77,896  & 82.3 & 138,042   & 75,885    \\ \bottomrule
\end{tabular}
\label{table:1}
\end{table}

\subsection{Verification of user classes}
The classification results of users in the previous section (i.e., CH, CB, NH, NB) affect the results of our analyses.
We confirmed the reliability of the classification using manual verification as described next. 

For bot classification, we randomly selected 200 accounts from our dataset, with 100 labeled as bots and 100 labeled as humans.
We provided two human workers with these account names and asked them to independently read profiles and tweets by each user and then classify the user as a bot or human. 
Then we checked the classification consistency by computing Cohen's kappa coefficient. 
The resulting $\kappa=0.68$ indicates substantial agreement between the two workers, indicating that the bot score threshold used is sufficiently reliable. 
Note that according to~\cite{10.2307/2529310}, Cohen's kappa value is interpreted as follows: 0.0-0.2 for slight agreement; 0.2-0.4 for fair agreement; 0.4-0.6 for moderate agreement; 0.6-0.8 for substantial agreement; and 0.8-1.0 for near perfect agreement.

A similar verification process was conducted for credible/non-credible classification with  two additional workers. 
They evaluated 100 randomly selected accounts, with 50 labeled as credible and 50 labeled as non-credible, by reading their tweets and profiles, and then classified them as credible or non-credible users. After that, Cohen's kappa was computed and resulted in $\kappa=0.70$, indicating a significant match. In this way, credible/non-credible classification based on the above-mentioned criteria is, likewise, sufficiently reliable.

\subsection{Retweet behavior analysis}
\label{subsec:method_rt_net}
To examine information spreading patterns by topic, we constructed a retweet network for each topic, in which nodes represent users and a directed edge was made from the source to the target if a target user is retweeted by a source user. 
The structural features were quantified using these retweet networks.
For visibility, we illustrated the retweet network by including only bots and selected accounts (as anchors) by using the software Gephi \cite{ICWSM09154}, with the graph layout algorithm ForceAtlas2 \cite{Jacomy2014}. 
We highlighted the selected users with a large indegree that includes famous politicians, well-known mainstream media, and right-wing media from the top 40 indegree users. 
Moreover, we compared temporal patterns of retweet activities by topic among four types of users (i.e, CH, CB, NH, NB).

Furthermore, we quantified how bots interacted with media and celebrities using the aggregated retweet network ($n=211$) based on the bots that appeared across the topics.
For this, we first identified 19 credible media and celebrity accounts (CM and CC), and 12 non-credible counterparts (NM and NC) in reference to their user profiles, their Wikipedia articles, and the official websites, Media Bias / Fact Check, \footnote{\url{https://mediabiasfactcheck.com}} etc.
Then, we classified media and celebrity accounts as: (a) those retweeted by non-credible bots only; (b) those retweeted by both credible and non-credible bots; and (c) those retweeted by credible bots only.
The network visualization process was the same as above. 

\subsection{Retweeted contents analysis}
\label{subsec:method_rt_content}
We also looked at differences in the contents of the retweeted URLs in each topic. 
Retweets in our COVID-19 dataset did not contain a sufficient amount of texts, instead using hyperlinks or URLs to online articles. 
Thus, we focused on tweets including URLs to online articles and collected these articles retweeted by credible/non-credible humans and bots, separately. 
We characterized the articles based on terms (nouns), with their importance measured by the TF-IDF score. For this analysis, we selected the top 30 terms from credible users and the top 30 terms from non-credible users, and then merged them without duplicates from the collected articles.

TF-IDF stands for term frequency--inverse document frequency and is commonly used in natural language processing (NLP). TF-IDF is calculated as follows:
\begin{equation}
TF \times IDF,
\end{equation}
where $TF$ is the number of a given term (noun). We used the following formula for IDF:
\begin{equation}
IDF=\ln \frac{N}{d},
\end{equation}
where $N$ represents the total number of documents, and $d$ represents the number of documents that include the term. 

To compare important terms used in articles retweeted by credible and non-credible users, we summarized TF-IDF values by using the laterality index ($LI$)~\cite{Sasahara2019a}, defined as follows:

\begin{equation}
LI=\frac{C-NC}{C+NC}, LI \in [-1,1],
\end{equation}
where $C$ is the TF-IDF score for terms used in articles retweeted by credible users and $NC$ is used for terms used in articles retweeted by non-credible users. $LI$ compares the importance of a term between credible sites and non-credible sites. 
A negative $LI$ indicates that the term is characteristic of non-credible sites; a positive $LI$ indicates that the term is characteristic of credible sites; $LI=0$ indicates that the term is equally important in both sites. 

\section{Results}
\subsection{Segregated structures of retweet networks}
\begin{figure}[H]
    \centering
    \begin{subfigure}[b]{0.5\textwidth}
        \centering
        \includegraphics[width=\textwidth]{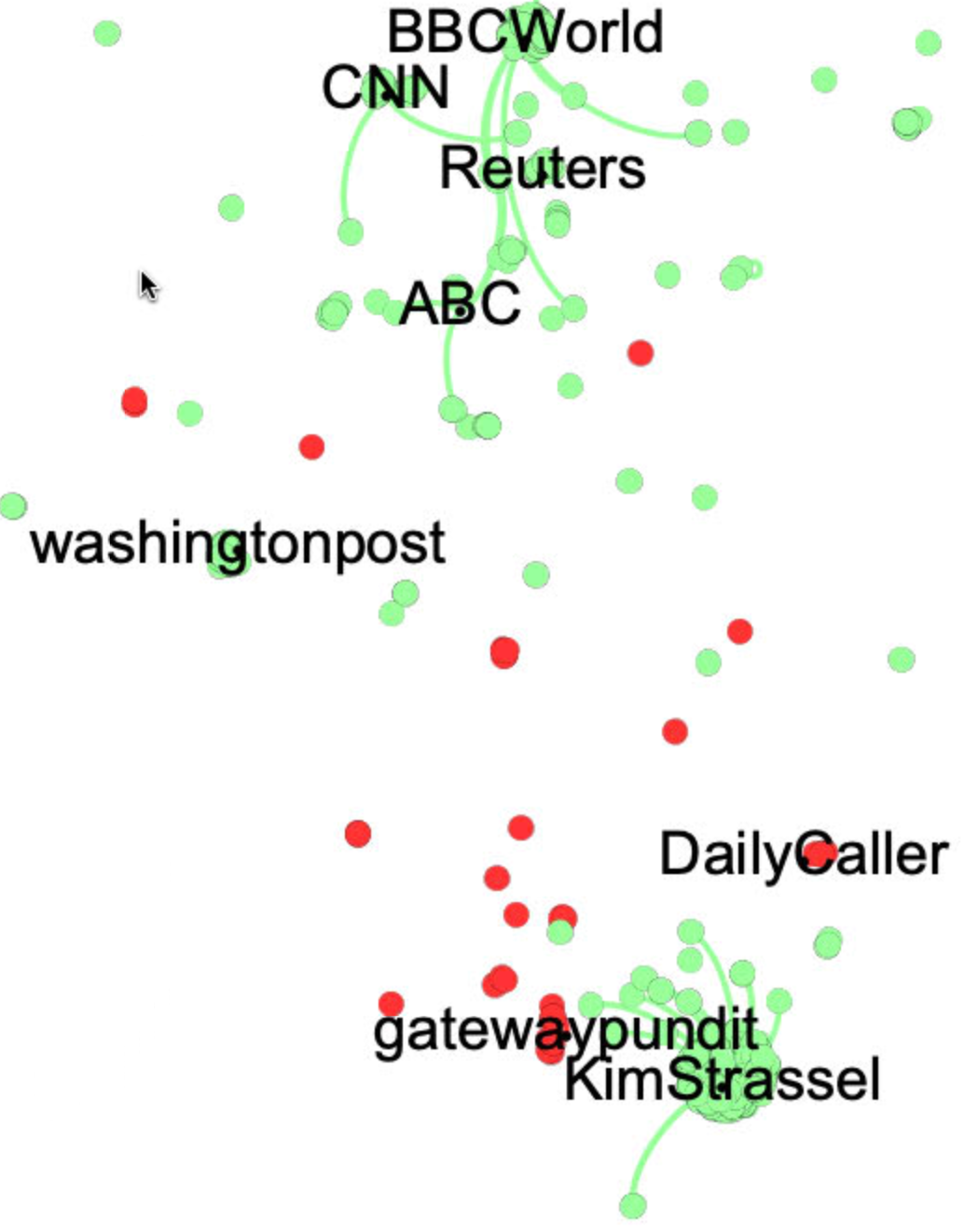}
        \caption[WHO]%
        {{\small WHO}}    
        \label{WHO}
    \end{subfigure}
    \hfill
    \begin{subfigure}[b]{1.0\textwidth}  
        \centering 
        \includegraphics[width=\textwidth]{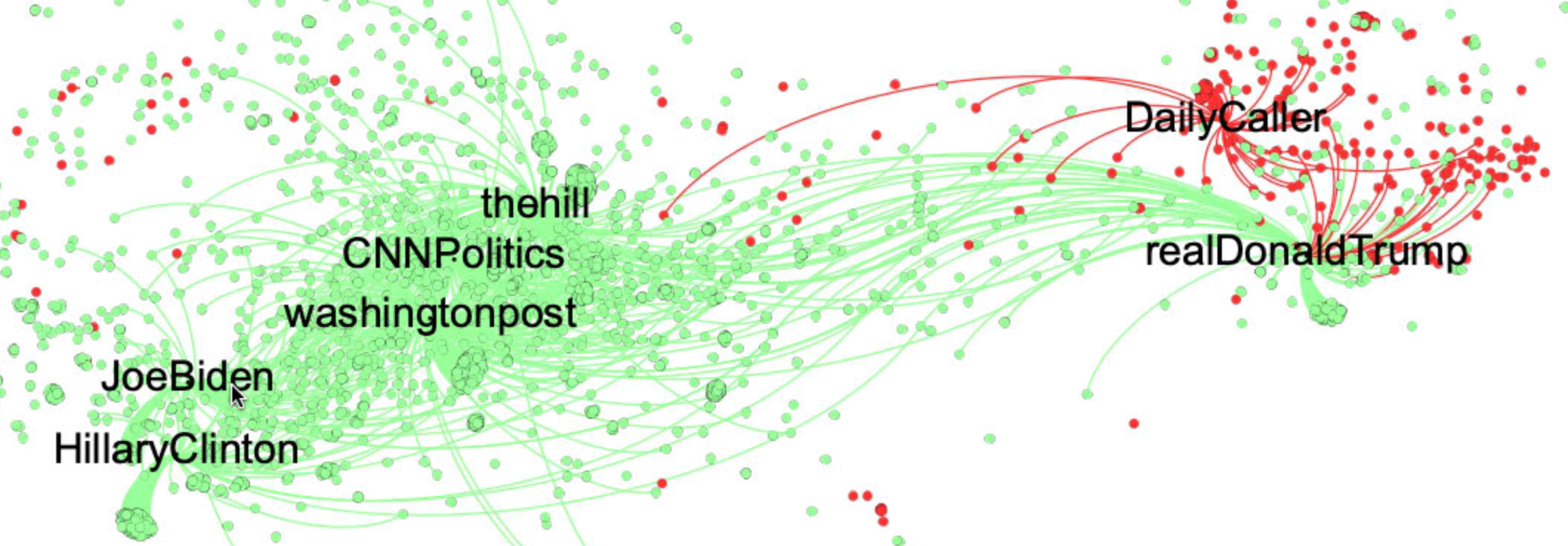}
        \caption[]%
        {{\small Trump}}    
        \label{trump}
    \end{subfigure}
    \vskip\baselineskip
    \begin{subfigure}[b]{0.5\textwidth}   
        \centering 
        \includegraphics[width=\textwidth]{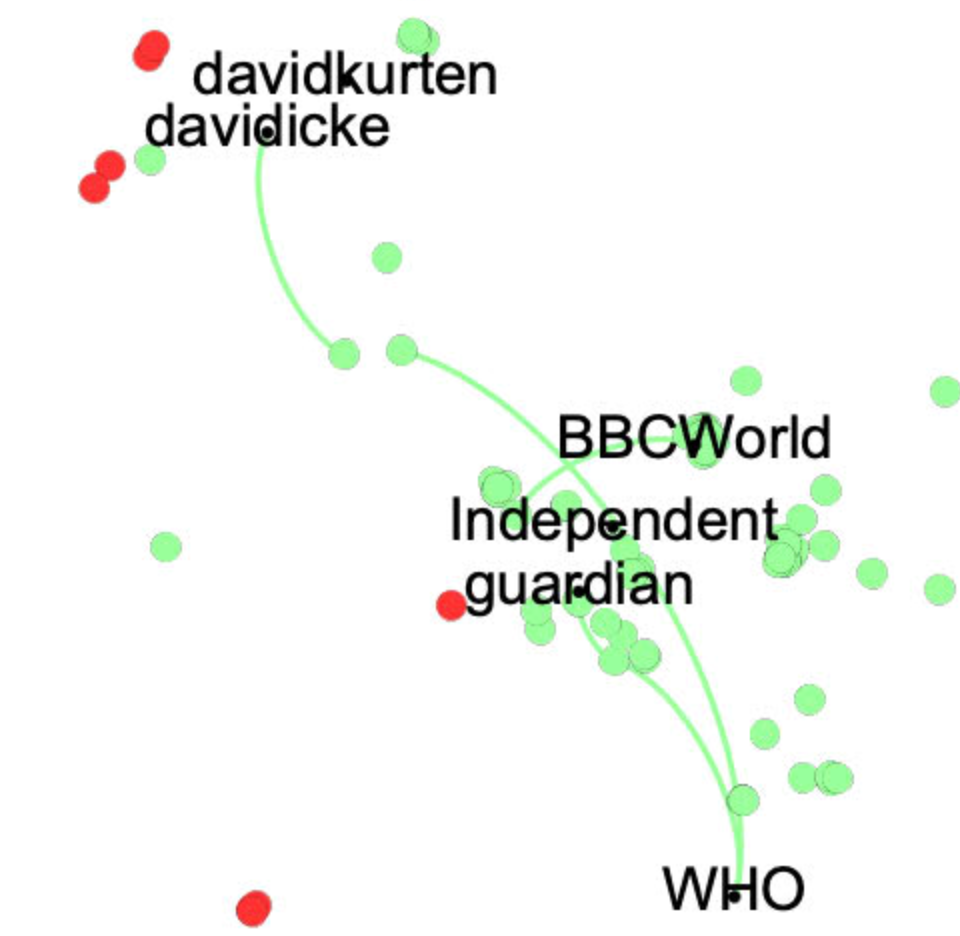}
        \caption[]%
        {{\small 5G}}    
        \label{5G}
    \end{subfigure}
    \hfill
    \begin{subfigure}[b]{0.9\textwidth}   
        \centering 
        \includegraphics[width=\textwidth]{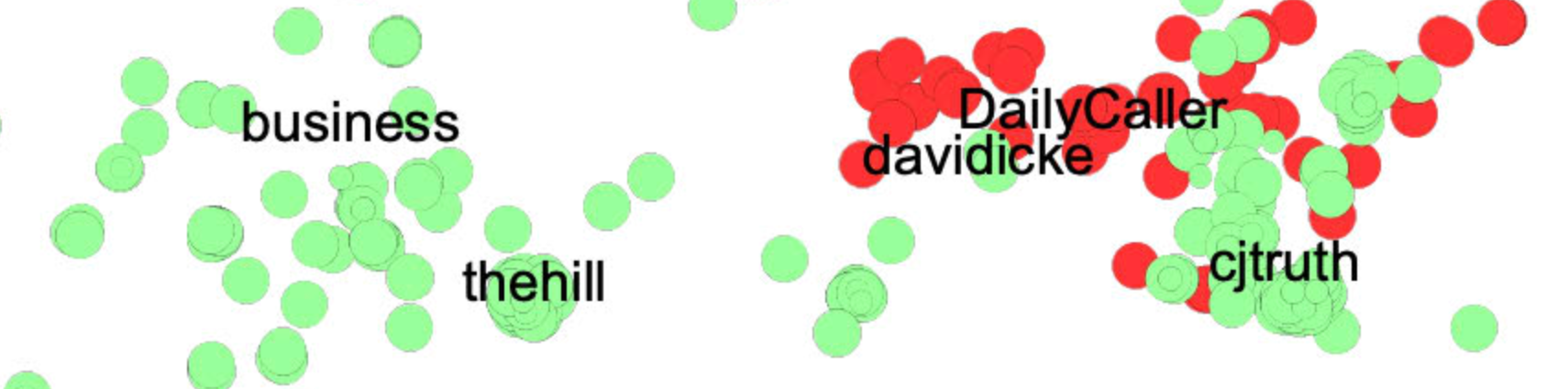}
        \caption[]%
        {{\small Bill Gates}}    
        \label{billgates}
    \end{subfigure}
    \caption{Retweet networks related to ``WHO'',``Trump'', ``5G'', and ``Bill Gates''. Red nodes indicate non-credible bots; green nodes indicate credible bots. Edges represent retweets among bots, and between bots and popular accounts (labelled).}
\label{fig:RT}
\end{figure}

We looked at the retweet interactions between humans and bots for each topic using the preprocessed COVID-19 tweets. The resulting retweet networks are shown in Fig.~\ref{fig:RT}. It is notable that segregated structures emerged in all the topics considered, with dense connections inside and sparse connections in between. 

In the ``WHO'' network ($n=88,719$), 3.8\% of labelled users are non-credible users while 19.7\% are credible bots.
The credible group contained official media accounts such as ``@washingtonpost'', ``@ABC'', ``@Reuters'', ``@CNN''and ``@BBCWorld'' and were separated from the non-credible group containing ``@DailyCaller'', ``@gatewaypundit'' and ``@KimStrassel''(Fig. \ref{WHO}). We found that non-credible bots were appearing around the US conservative columnist ``KimStrassel'' (Kimberley Strassel) as well as ``@DailyCaller'' (an American right-wing misinformation website) and ``@gatewaypundit'' (an American far-right website publishing misleading news).
These results imply that the non-credible bots might be attempting to interact with politically right-leaning users to increase these users' exposures to negative information. 
Although the WHO itself is a neutral topic, partisan asymmetry was visible during the COVID-19 infodemic.

Previous research has found that the retweet network of the 2010 US midterm elections showed typical ``left'' and ``right'' segregated groups \cite{Conover2012}. Thus, we examined whether the ``Trump'' retweet network shares the similar features. Fig. \ref{trump} shows the Trump network ($n=1,125,251$) where 3.2\% of the labelled users are non-credible bots while 23.5\% are credible bots. 
Here ``@HillaryClinton'' (Hillary Clinton) and ``@JoeBiden'' (Joe Biden), representing the progressives clustered together, were distant from the conservative cluster with ``@realDonaldTrump'' (Donald Trump). The political echo chamber was reobserved in 2020 in the context of the COVID-19 infodemic.
A notable finding is that ``@realDonaldTrump'' was mostly retweeted by non-credible bots (shown in red), whereas ``@HillaryClinton'' and ``@JoeBiden'' were less so. 

As far as ``5G'' is concerned, two separate groups were observed again in the retweet network ($n=67,523$), in which 1.62\% of the labelled users are non-credible bots and 8.82\% are credible bots (Fig. \ref{5G}).
One side of the network includes ``@davidicke'' (David Icke) and ``@davidkurten'' (David Kurten). The former is reported to be a conspiracy theorist, and the latter is currently a member of the UK Independence Party (right-wing populist party) and has been since 2016 \cite{BBCUKIP,Skynews,UropeanConservative}. They were the two most-retweeted users in the 5G conspiracy topic. By contrast, the mainstream British media ``@BBCWorld'' and ``@WHO'' were located on the other side of the network in Fig. \ref{5G}. More non-credible bots were involved on the side of ``@davidicke'', while there were more credible bots on the other side. Although ``5G'' was considered as a popular conspiracy theory in the early days of COVID-19 pandemic, a larger number of non-credible bots was not observed in comparison with other topics. 

``Bill Gates'' is another conspiracy theory topic as mentioned earlier. In regard to this topic, 5.95\% of the labelled users in the retweet network ($n=94,584$) are non-credible bots while 18.0\% are credible bots (Fig. \ref{billgates}). Again, we can find the ``@davidicke'' and ``@DailyCaller'', which are occupied by non-credible bots, on the right side, while a cluster on the left is covered by credible bots with legitimate media including ``@business'' and ``@thehill''. 
Moreover, ``@davidicke'' was observed in both ``5G'' and ``Bill Gates'' conspiracy topics; this account was suspended by Twitter and is no longer accessible. 
It can be observed that there are no visible links among these bots in the ``Bill Gates'' network. 
This is because these bots and labelled accounts did not have mutual retweets; for example,  ``@DailyCaller'' was retweeted by 336 humans, 294 of whom were non-credible humans.

\begin{figure}[t]
  \centering
  \includegraphics[width=\textwidth]{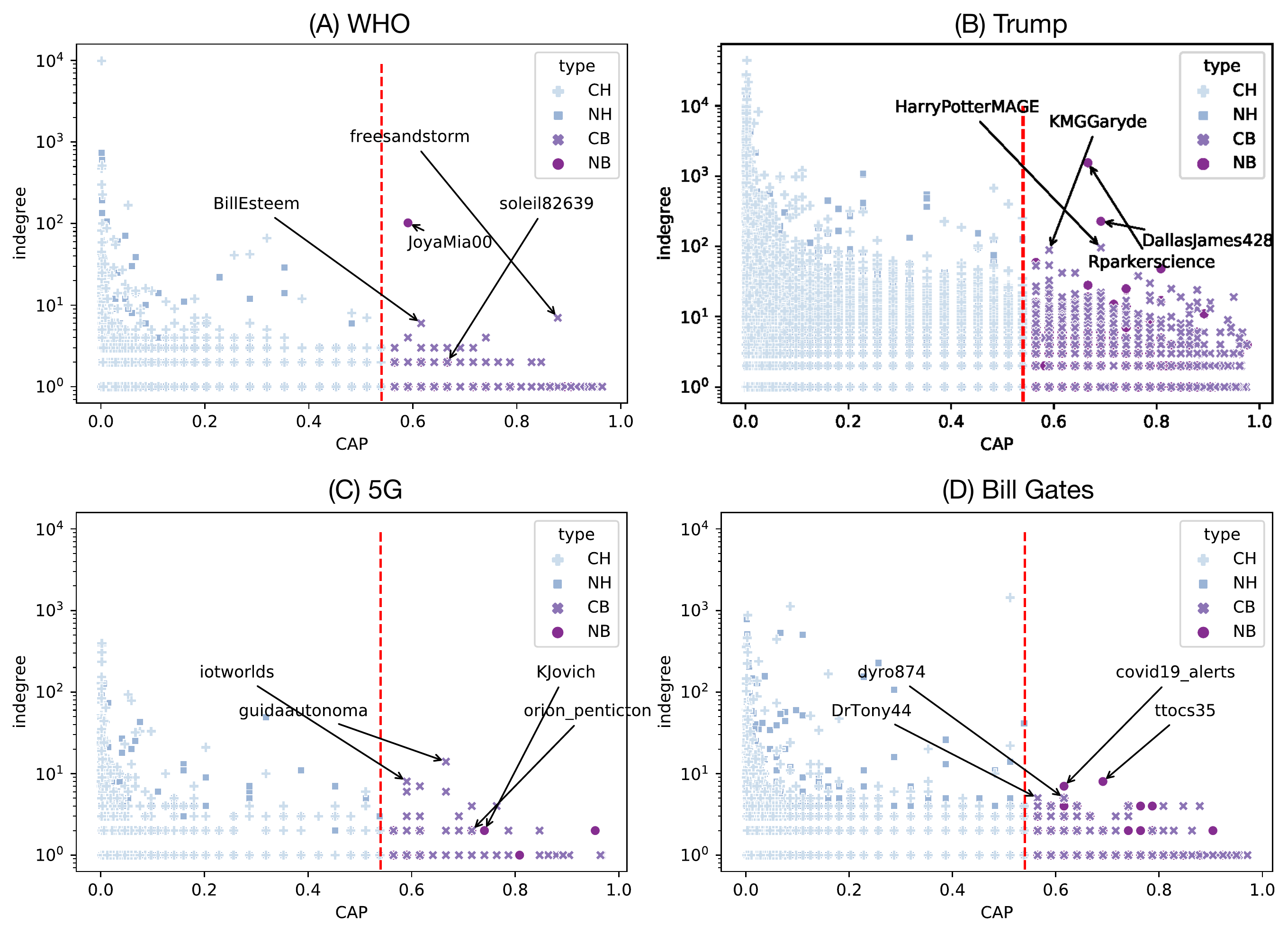}
  \caption{Degrees vs. Bot Score (CAP) in ``WHO'', ``Trump'', ``5G" and ``Bill Gates'' topics. The red dashed line is the threshold for bots/humans classification (CAP=0.54).}
  \label{fig:botometer_vs_indegree}
\end{figure}

Then, we quantified indegrees (the numbers of retweeted posts by different users, used as a measure for engagement) as a function of the bot score (CAP). 
The resulting scatter plots are shown in Fig.~\ref{fig:botometer_vs_indegree}. The complementary cumulative distribution Function (CCDF) of indegrees is shown in the supplementary information (SI). 
The majority of users are obviously credible humans. 
It turns out that the indegrees tend to be inversely proportional to the bot score and, on average, indegrees for humans are larger than those for bots in all the topics. 
Compared with humans, bots were, therefore less engaging in retweets in general. 
However, average indegrees of non-credible bots are higher than those of credible bots ($t$-test, $p=0.00047$). 

There are several outliers of non-credible bots that have larger indegrees in Fig. \ref{fig:botometer_vs_indegree}. 
For example, `` @JoyaMia00''(NB) is an MAGA (Make America Great Again) user and a Trump supporter; ``@soleil82639'' (NB) retweets posts in at least three languages including English, Japanese, and French; These posts are related to history and politics as well as COVID-19. In addition, ``@freesandstorm'' (CB) is always retweeting climate and environment-related posts (``@BillEsteem'' (CB) is closed.) (Fig.  \ref{fig:botometer_vs_indegree}a). In the Trump topic, ``@DallasJames428'' (NB) and ``@Rparkerscience'' (NB) are suspended. ``@HarryPotterMAGE'' (CB) is retweeting political tweets while ``@KMGGaryde'' (CB) is closed (Fig. \ref{fig:botometer_vs_indegree}b); In the ``5G'' topic, ``@iotworlds'' (CB) is closed. From the page of ``@guidaautonoma'' (CB) we can say that the users is enthusiastic about autonomous car; ``@orion\_pentict'' (NB) retweets many anti-Biden posts, while ``@KJovic'' is an anti-lockdown supporter (Fig. \ref{fig:botometer_vs_indegree}c). In the Bill Gates topic,  ``@dyro874'' (CB) is suspended; ``@DrTony44'' is now retweeting posts not related to COVID-19 but to entertainment; ``@covid19\_alert'' (NB)
is retweeting COVID-19 posts; and ``@ttocs35'' is now suspended (Fig. \ref{fig:botometer_vs_indegree}d). 
These non-credible bots were actually retweeted as many times as those of human accounts and, thus, were as equally influential as humans. 
We confirmed that some outlier bots were actually suspended by Twitter and are no longer active.
These results indicate that, although the number of retweets by non-credible humans could be larger than the number by non-credible bots, the effects of the latter are still non-negligible because of the existence of the outliers as well as the parasitic nature of non-credible bots, which we will discuss later.

Overall, two segregated networks of information-spreading emerged in all the topics considered. It turns out that one side of the dense connected components was propagating credible sources mostly by credible bots. By contrast, the other component was diffusing non-credible information, mainly amplified by non-credible bots. 

\subsection{Parasitic natures of bots}
\begin{figure}[H]
\centering
\begin{subfigure}[b]{0.95\textwidth}
\centering
\includegraphics[width=1.0\textwidth]{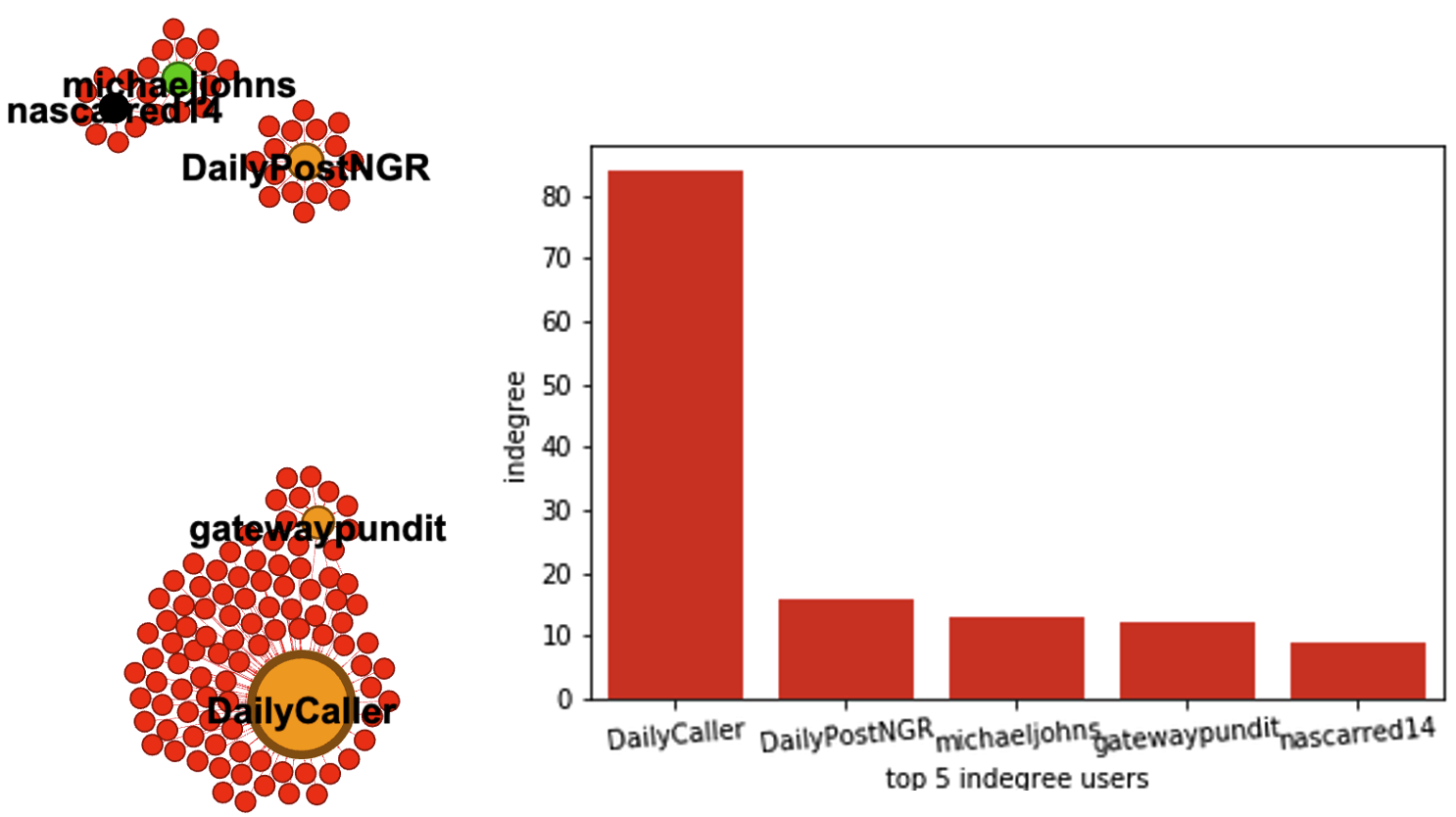}
\caption{}
\label{fig3:NBonly}
\end{subfigure}

\begin{subfigure}[b]{1\textwidth}
\centering
\includegraphics[width=0.95\textwidth]{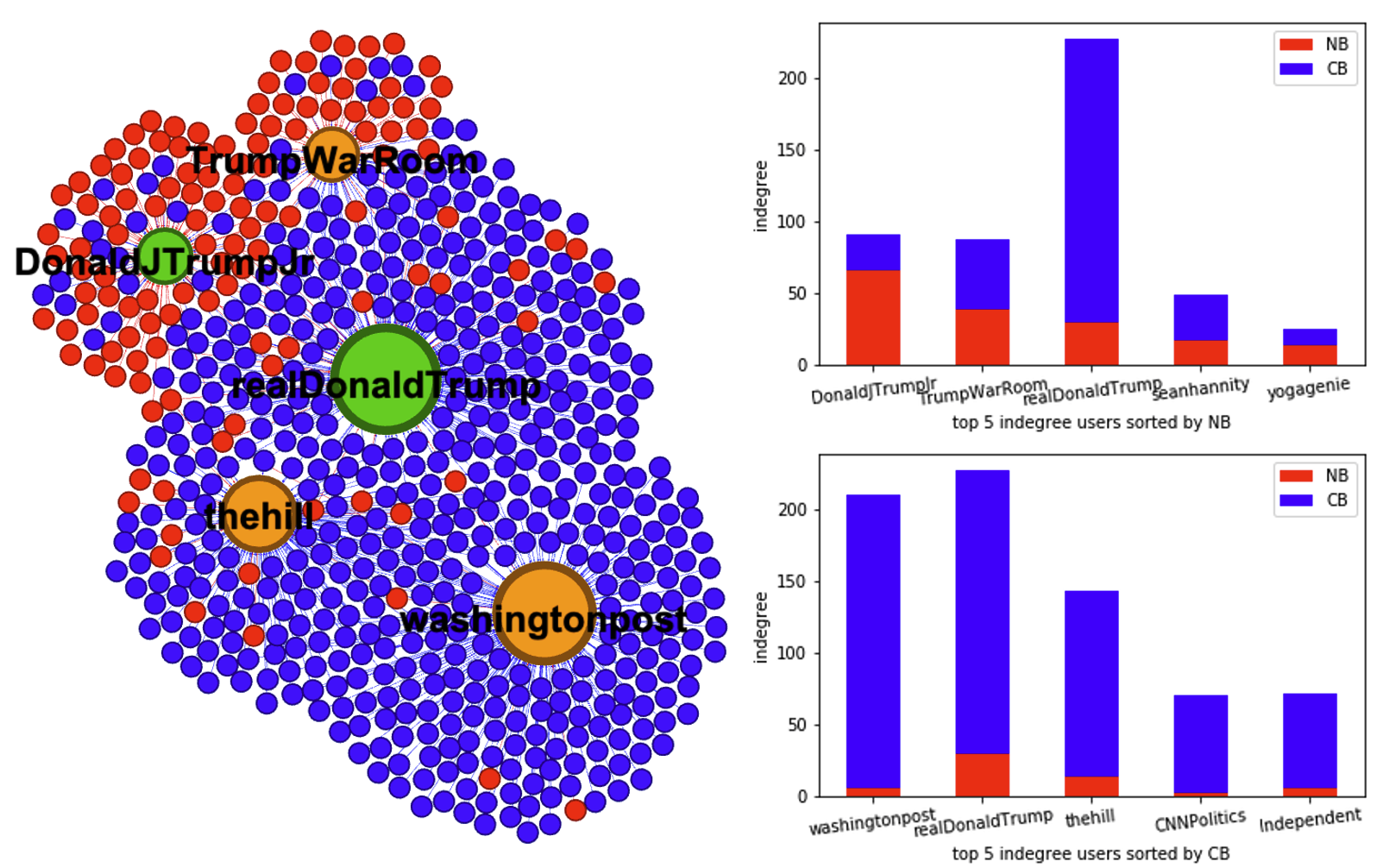}
\caption{}
\label{fig3:NB_CB}
\end{subfigure}

\begin{subfigure}[b]{1\textwidth}
\centering
\includegraphics[width=0.95\textwidth]{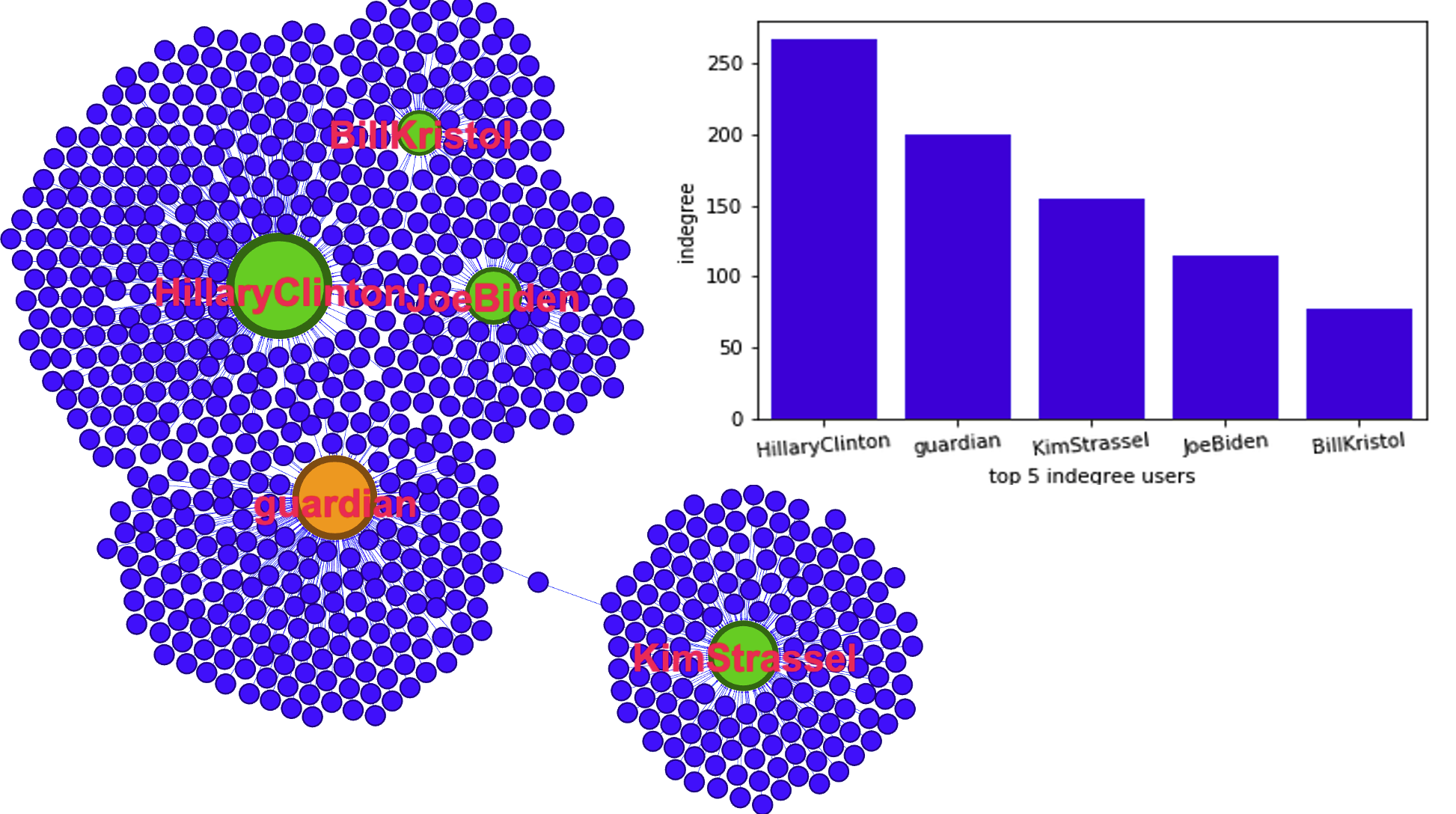}
\caption{}
\label{fig3:CBonly}
\end{subfigure}
\caption{Celebrities and media accounts retweeted by (a) non-credible bots only; (b) both credible and non-credible bots; (c) credible bots only. Green denotes celebrity and orange denotes media. Accounts with black labels were all suspended by Twitter (as of Aug. 7, 2021). The size of a node is proportional to its indegee.}
\label{fig3}
\end{figure}

Next, we examined how bots selectively amplified voices from media and celebrity accounts in the aggregated retweet networks mentioned in Sec.~\ref{subsec:method_rt_net}.
Fig.~\ref{fig3} shows the results by bot type.
The top-five most-retweeted accounts were plotted alongside. 

In Fig.~\ref{fig3:NBonly}, we can clearly see that non-credible bots are parasitic on far-right mass media accounts such as ``@DailyCaller''($k_{in}=84$) in terms of their retweets, although the indegrees of the other four are much smaller. Note that ``@DailyPostNGR''is a Nigerian local newspaper; ``@michaeljohns'' is an American conservative commentator, speechwriter, and policy analyst; also, ``@nascarred14'' is now suspended. 

Looking at the accounts retweeted by both credible and non-credible bots in Fig.\ref{fig3:NB_CB}, it turns out that most non-credible bots are also parasitic on right-leaning celebrity and media accounts, including ``@DonaldJTrumpJr'' ($k_{in}=61$). 
His most popular article is reposted from the ``DailyCaller'' website, saying that ``Chinese Government Finally Acknowledges Underreporting Coronavirus Cases''. 
Other accounts include ``@TrumpWarRoom'' (``The office of Donald J Trump'') \footnote{\url{https://https://donaldjtrump.com}}, ``@seanhannity''(an American talk show program presenter and conservative political commentator.) and ``@yogagenie'' (suspended).
By contrast, most credible bots are preferentially amplified voices from legitimate media accounts, such as ``@washingtonpost'' ($k_{in}=205$). Other accounts include ``@thehill'', ``@CNNPolitics'' and ``@Independent''.
A marked exception is ``@realDonaldTrump'' whose posts were largely shared not only by non-credible bots ($k_{in}=29$) but also credible bots ($k_{in}=199$). 

Fig.\ref{fig3:CBonly} shows that major celebrities and media were selectively retweeted only by credible bots, regardless of their political attitude. For instance, ``@Kimstrassel'' (American conservative writer, $k_{in}=155$) and ``@HillaryClinton'' (US politician, $k_{in}=267$). The other accounts include ``@guardian'' (British newspaper), ``@JoeBiden'' and ``@BillKristol'' (American neoconservative political analyst).

These results further support that non-credible bots cannot be ignored in the context of the COVID-19 infodemic due to their parasitic natures toward popular right-leaning accounts. 

\subsection{Temporal patterns of retweets in humans and bots}
\begin{figure}[H]
  \centering
  \includegraphics[width=\textwidth]{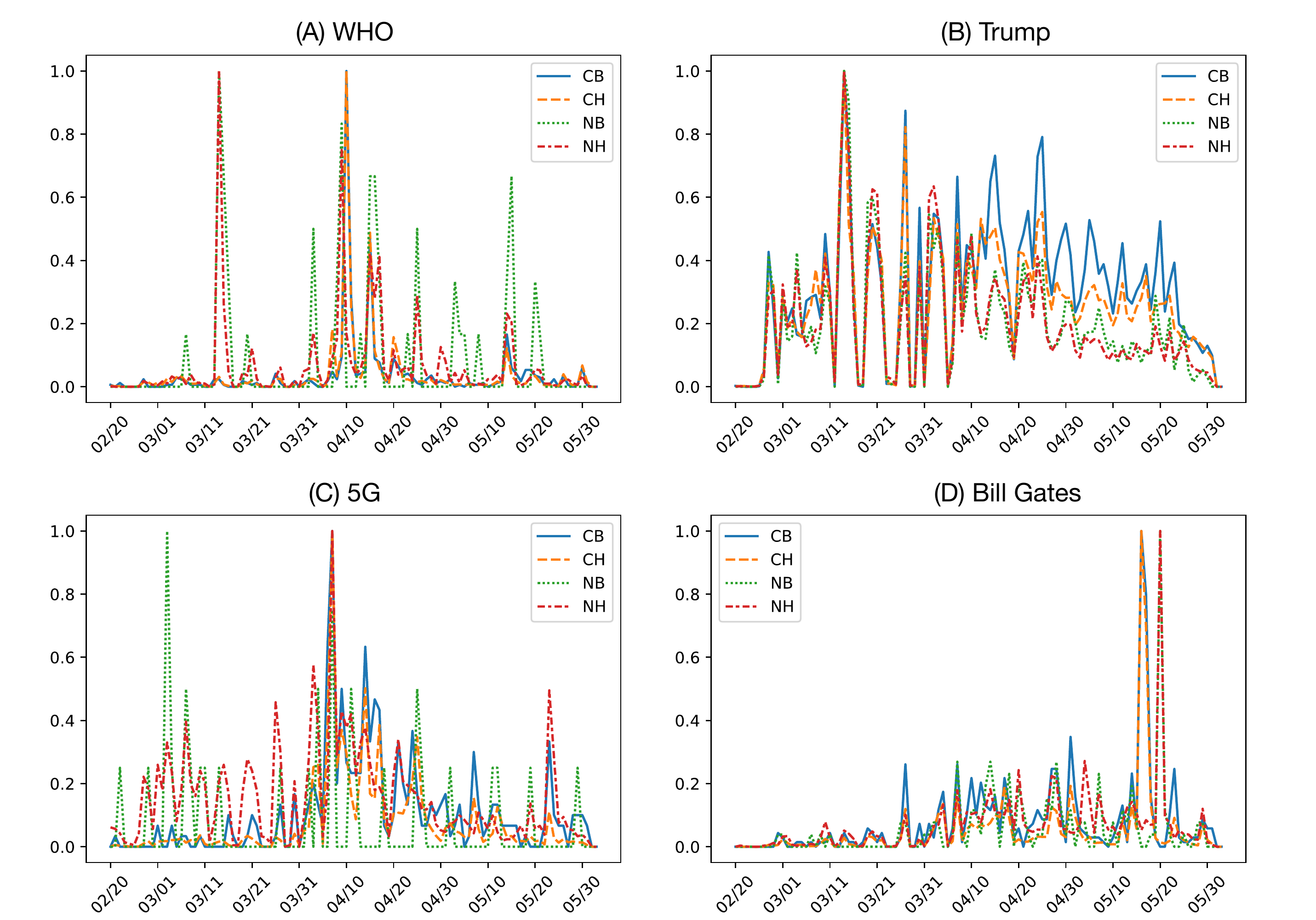}
  \caption{Retweet count series (scaled 0-1) generated for bots and humans in four topics.}
  \label{fig:4topics}
\end{figure}

We assumed that non-credible bots' behaviors are correlated with non-credible humans rather than credible accounts, because the intention of non-credible bots would be to amplify the spread of misinformation including conspiracy theories. Thus, we quantified temporal patterns of retweet behaviors in humans and bots. 
For comparisons among credible/non-credible humans and bots, we scaled daily retweet counts between 0 and 1, respectively.
Fig. \ref{fig:4topics} shows daily retweet series by humans and bots for each topic, in which the patterns of retweets increase following similar trends.

To statistically confirm this observation, we measured the Pearson correlation coefficient of temporal oscillations of retweets generated by these users. 
The results are summarized in Table~\ref{Pearson}.
This reveals that retweet actions by non-credible bots correlated with those by non-credible humans to a much higher degree than by credible humans in all of the topics.
The above assumption is therefore partially supported. 
We further consider this assumption in the next section by looking at commonality in retweets generated by bots and humans.

\begin{table}[t]
\centering
\caption{Pearson correlation coefficients of retweets between NB and NH and between NB and CH.}
\label{Pearson}
\begin{tabular}{cccc}
\toprule
Topic                       & Type     & Correlation Coefficient & P-value     \\ \midrule
\multirow{2}{*}{WHO}        & NB \& NH & 0.84                    & 6.05E-29    \\
                            & NB \& CH & 0.15                    & 0.1250527   \\ \hline
\multirow{2}{*}{Trump}      & NB \& NH & 0.96                    & 4.65E-61    \\
                            & NB \& CH & 0.82                    & 1.81E-26    \\ \hline
\multirow{2}{*}{5G}         & NB \& NH & 0.45                    & 1.31E-06    \\
                            & NB \& CH & 0.32                    & 0.001031208 \\ \hline
\multirow{2}{*}{Bill Gates} & NB \& NH & 0.91                    & 1.46E-41    \\
                            & NB \& CH & 0.04                    & 0.712220673 \\ \bottomrule
\end{tabular}
\end{table}

\subsection{Commonality in retweets by humans and bots}
\begin{figure}[H]
\centering
\begin{subfigure}[b]{0.95\textwidth}
\centering
\includegraphics[width=1.2\textwidth]{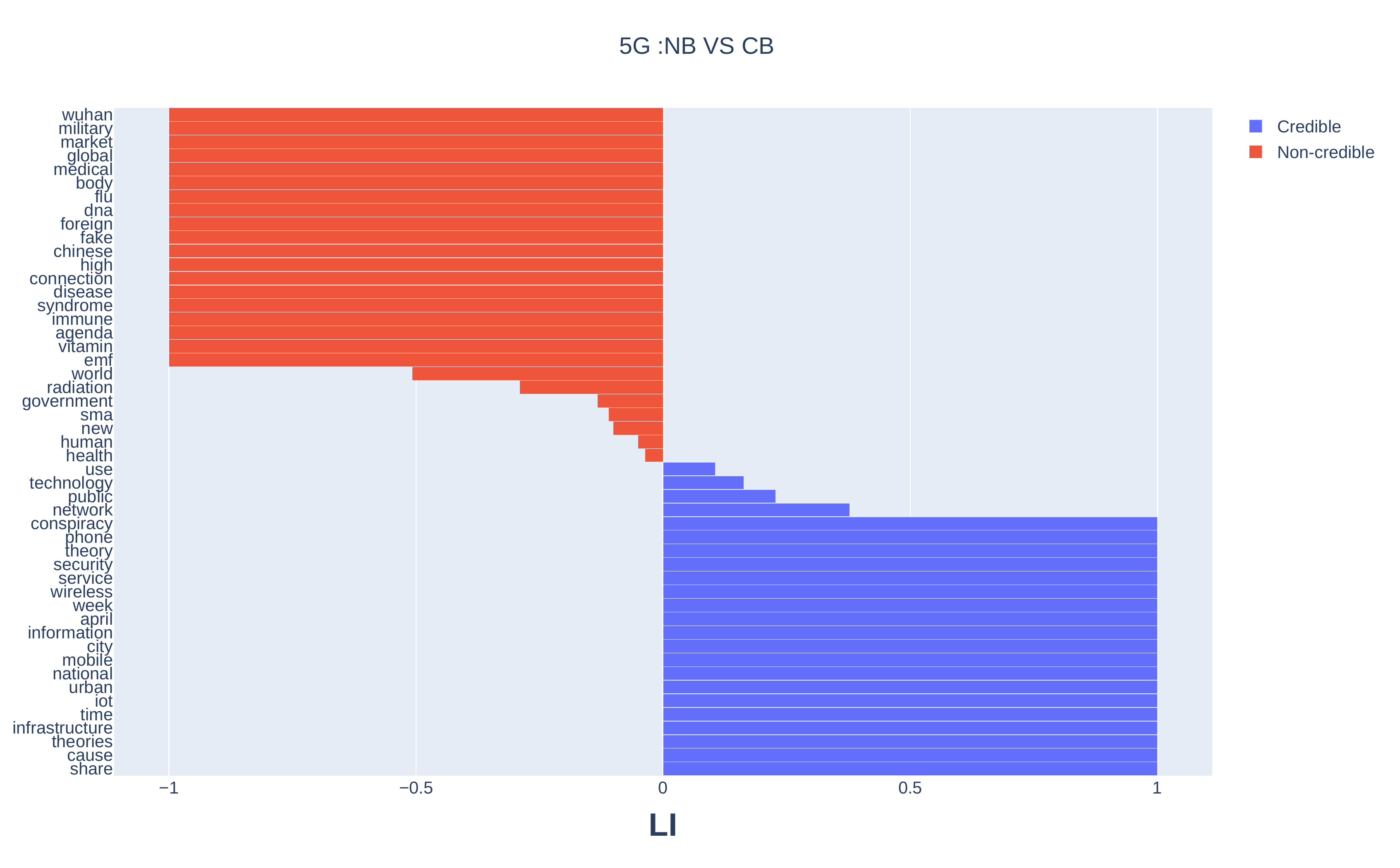}
\caption{}
\label{fig:tfidf1}
\end{subfigure}

\begin{subfigure}[b]{1\textwidth}
\centering
\includegraphics[width=1.15\textwidth]{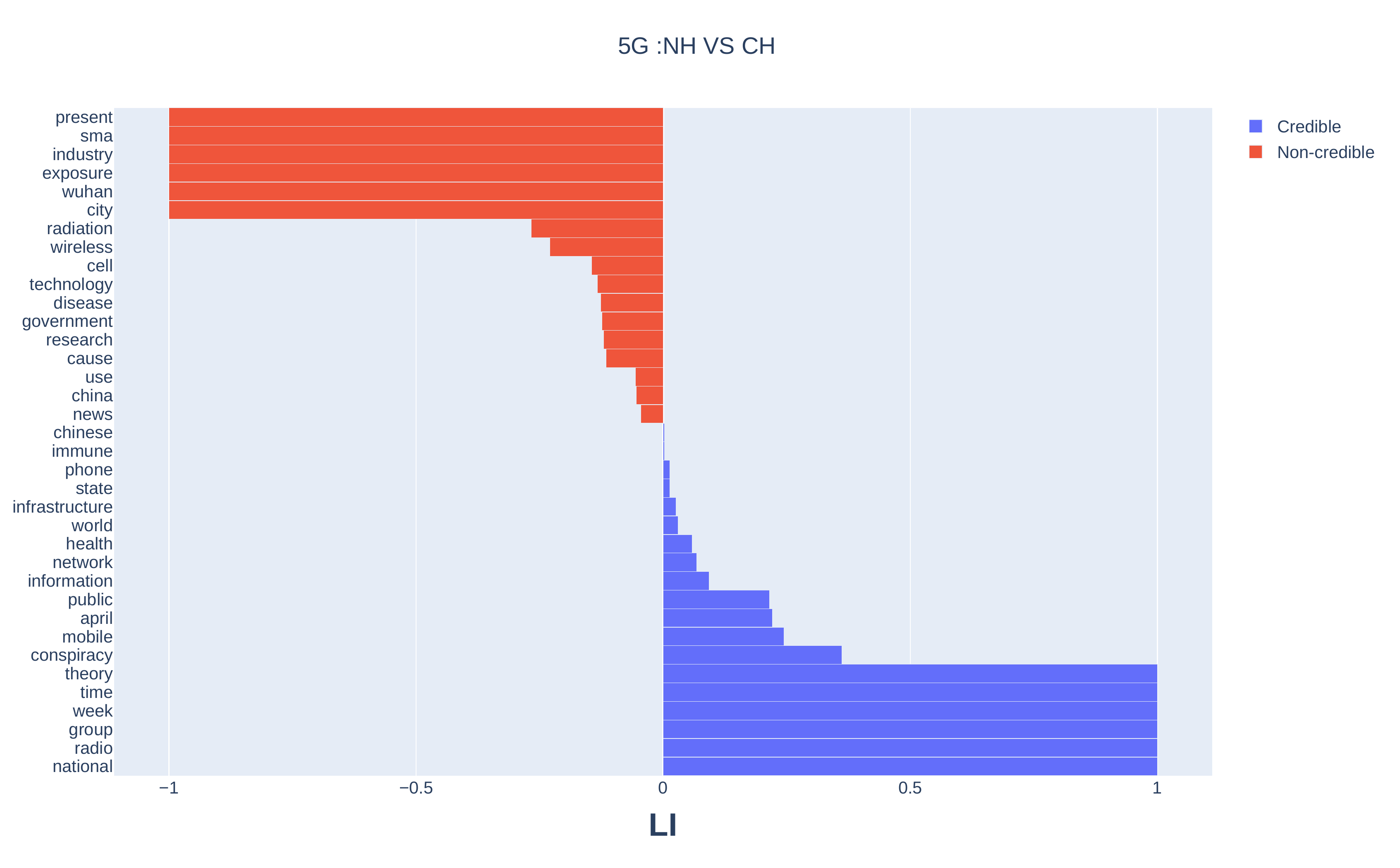}
\caption{}
\label{fig:tfidf2}
\end{subfigure}
\caption{Term importance in retweeted articles in the ``5G'' topic: (a) non-credible bots vs. credible bots; (b) non-credible humans vs. credible humans.  Red bars indicate word importance for non-credible users, whereas blue bars indicate credible users.}
\label{fig:tfidf}
\end{figure}

Finally, we examined terms (nouns), domains (URLs), and users that commonly appeared in retweets generated by humans and bots. 
Consider the 5G topic, for example; Fig. \ref{fig:tfidf} compares term importance (measured by TF-IDF) on the 5G-related articles retweeted by bots and humans.
As seen in the red bars of Fig. \ref{fig:tfidf} (a) and (b), the non-credible bots selectively retweeted articles that include China-related terms such as ``wuhan'', ``china'', ``chinese'', and the non-credible humans had the same tendency. Sample articles include the following:

\vspace*{1\baselineskip}
\emph{``The China Coronavirus 5G Connection is a very important factor when trying to comprehend the coronavirus (formerly abbreviated 2019-nCoV, now COVID-19) outbreak.''}~\footnote{\url{https://worldtruth.tv}}
\vspace*{1\baselineskip}

Such articles aimed to connect ``china'' with coronavirus subjectively and highlighted that Wuhan was a test city chosen for China's 5G rollout.
By contrast, the blue bars in Fig.~\ref{fig:tfidf} (a) and (b) show that the credible bots and humans retweeted articles that include the word ``conspiracy.''
This suggests that both credible bots and humans tended to retweet articles designed to  warn the readers about the conspiracy theory. Sample articles include:

\vspace*{1\baselineskip}
\emph{``One theory claims that the novel coronavirus originated in Wuhan because the Chinese city had recently been rolling out 5G. It’s now supposedly spread to other cities that are also using 5G. These false conspiracy theories neglect to mention that a highly contagious virus would naturally spread more in densely populated cities with access to 5G, and that the coronavirus pandemic has hit countries like Iran and Japan where 5G isn’t in use yet.''}~\footnote{\url{https://www.theverge.com}}
\vspace*{1\baselineskip}

Thus, the 5G conspiracy theory went viral inside of an echo chamber consisting of non-credible accounts while shutting down contrasting voices from a credible cluster.

The same linguistic analysis was conducted for the ``WHO'', ``Trump'', and ``Bill Gates'' topics, and resulted in the similar linguistic features (see Figs. 2a-f in SI). 
That is, the articles including China-related terms were preferentially shared by both non-credible bots and humans in the Trump topic and others.
These consistent patterns suggest that non-credible humans were diffusing China-related mis/disinformation and conspiracy theories, amplified by non-credible bots with political inclinations.

We also calculated the shared terms ratio in each topic. 
The non-credible bots and humans shared 57\%, 90\%, 50\%, and 30\% of terms in the retweeted articles for ``WHO'', ``Trump'', ``5G'', and ``Bill Gates'' topics, respectively (cf. credible bots and humans shared 73\%, 93\%, 70\%, and 40\% of terms, respectively).
This indicates that the non-credible bots and humans are topically similar, and therefore, one may not be able to easily distinguish humans or bots based on the observation of a few tweets.
These results, again, suggest that the non-credible humans were diffusing China-related misinformation and misleading claims, and that the non-credible bots were probably amplifying their effects. 
We will discuss the implications of this later. 


\begin{table}[H]
\caption{Popular retweeted domains and users in the 5G topic. (a) Top 10 domains retweeted by credible humans (CH), credible bots (CB), non-credible humans (NH), and non-credible bots (NB). (b) Top 10 users retweeted by CH, CB, NH, and NB. Green denotes credible domains or users; red denotes non-credible domains or users; blue denotes unknown domains or users.}
\label{table:5G_common}
\begin{subtable}[t]{0.48\textwidth}
\caption{Top 10 domains retweeted by}
\begin{tabular}{cllll}
\toprule
\multicolumn{1}{l}{} &
  \multicolumn{1}{c}{\textbf{NB}} &
  \multicolumn{1}{c}{\textbf{CB}} &
  \multicolumn{1}{c}{\textbf{NH}} &
  \multicolumn{1}{c}{\textbf{CH}} \\ \hline
\textbf{1} &
  {\color{red} beforeitsnews.com} &
  {\color{ForestGreen} bbc.co.uk} &
  {\color{red} worldtruth.tv} &
  {\color{ForestGreen} theguardian.com} \\ \hline
\textbf{2} &
  {\color{red} dailypost.ng} &
  {\color{ForestGreen} theguardian.com} &
  {\color{red} express.co.uk} &
  {\color{ForestGreen} bbc.co.uk} \\ \hline
\textbf{3} &
  {\color{red} worldtruth.tv} &
  {\color{ForestGreen} bbc.com} &
  {\color{red} infowars.com} &
  {\color{ForestGreen} bbc.com} \\ \hline
\textbf{4} &
  {\color{red} zerohedge.com} &
  {\color{ForestGreen} reuters.com} &
  {\color{red} beforeitsnews.com} &
  {\color{ForestGreen} theverge.com} \\ \hline
\textbf{5} &
  {\color{blue} dailyrecord.co.uk} &
  {\color{ForestGreen} theverge.com} &
  {\color{red} humansarefree.com} &
  {\color{ForestGreen} businessinsider.com} \\ \hline
\textbf{6} &
  {\color{blue} today.ng} &
  {\color{ForestGreen} cnn.com} &
  {\color{red} neonnettle.com} &
  {\color{ForestGreen} cnn.com} \\ \hline
\textbf{7} &
  {\color{red} infowars.com} &
  {\color{ForestGreen} businessinsider.com} &
  {\color{red} thelastamericanvagabond.com} &
  {\color{ForestGreen} reuters.com} \\ \hline
\textbf{8} &
  {\color{blue} banned.video} &
  {\color{ForestGreen} nytimes.com} &
  {\color{red} dailycaller.com} &
  {\color{ForestGreen} ft.com} \\ \hline
\textbf{9} &
  {\color{blue} thetruthaboutcancer.com} &
  {\color{ForestGreen} newsweek.com} &
  {\color{red} paulcraigroberts.org} &
  {\color{ForestGreen} vox.com} \\ \hline
\textbf{10} &
  {\color{blue} rt.com} &
  {\color{ForestGreen} vox.com} &
  {\color{blue} thesun.co.uk} &
  {\color{ForestGreen} nytimes.com} \\ \bottomrule
\end{tabular}
\end{subtable}
\vfill

\begin{subtable}[t]{0.48\textwidth}
\caption{Top 10 users retweeted by}
\begin{tabular}{cllll}
\toprule
\multicolumn{1}{l}{} &
  \multicolumn{1}{c}{\textbf{NB}} &
  \multicolumn{1}{c}{\textbf{CB}} &
  \multicolumn{1}{c}{\textbf{NH}} &
  \multicolumn{1}{c}{\textbf{CH}} \\ \midrule
\textbf{1} &
  {\color{red} @shinethelight17} &
  {\color{ForestGreen} @Reuters} &
  {\color{red} @WorldTruthTV} &
  {\color{ForestGreen} @guardian} \\ \hline
\textbf{2} &
  {\color{red} @DailyPostNGR} &
  {\color{ForestGreen} @guidaautonoma} &
  {\color{red} @BILDERBERG\_GP} &
  {\color{ForestGreen} @rooshv} \\ \hline
\textbf{3} &
  {\color{red} @Laurel700} &
  {\color{ForestGreen} @Exchange5g} &
  {\color{blue} @davidicke} &
  {\color{ForestGreen} @guardiannews} \\ \hline
\textbf{4} &
  {\color{blue} @davidicke} &
  {\color{ForestGreen} @rooshv} &
  {\color{red} @shinethelight17} &
  {\color{ForestGreen} @verge} \\ \hline
\textbf{5} &
  {\color{red} @freezerohedge} &
  {\color{blue} @HaroldSinnott} &
  {\color{red} @TornadoNewsLink} &
  {\color{ForestGreen} @Omojuwa} \\ \hline
\textbf{6} &
  {\color{blue} @NigeriaNewsdesk} &
  {\color{ForestGreen} @verge} &
  {\color{red} @boblister\_poole} &
  {\color{ForestGreen} @Reuters} \\ \hline
\textbf{7} &
  {\color{blue} @BANNEDdotVIDEO} &
  {\color{blue} @ipfconline1} &
  {\color{red} @DailyPostNGR} &
  {\color{blue} @davidicke} \\ \hline
\textbf{8} &
  {\color{red} @WorldTruthTV} &
  {\color{blue} @Shirastweet} &
  {\color{blue} @BANNEDdotVIDEO} &
  {\color{ForestGreen} @Exchange5g} \\ \hline
\textbf{9} &
  {\color{blue} @owhy3} &
  {\color{ForestGreen} @nuskitconsultan} &
  {\color{blue} @davidkurten} &
  {\color{blue} @davidkurten} \\ \hline
\textbf{10} &
  {\color{blue} } &
  {\color{ForestGreen} @guardian} &
  {\color{red} @buttscornershop} &
  {\color{ForestGreen} @ruskin147} \\ \bottomrule
\end{tabular}
\end{subtable}
\end{table}

\begin{table}[H]
\centering
\caption{Common domains and users retweeted by non-credible users (NB and NH) and by credible users (CB and CH)}
\label{tab:top15summary}
\begin{subtable}[t]{0.48\textwidth}
\caption{Domains}
\label{tab:top15domains}
\begin{tabular}{ccc}
\toprule
\textbf{Topic}               & \textbf{NB $\cap$ NH} & \textbf{CB $\cap$ CH} \\ \midrule
\textbf{WHO}        & 6                 & 8                 \\ \hline
\textbf{Trump}      & 10                & 8                 \\ \hline
\textbf{5G}         & 3                 & 9                 \\ \hline
\textbf{Bill Gates} & 6                 & 8                 \\ \bottomrule

\end{tabular}
\end{subtable}
\vfill
\begin{subtable}[t]{0.48\textwidth}
\label{tab:top15users}
\caption{Users}
\begin{tabular}{ccc}
\toprule
\textbf{Topic}      & \textbf{NB $\cap$ NH} & \textbf{CB $\cap$ CH} \\ \midrule
\textbf{WHO}        & 8                 & 6                 \\ \hline
\textbf{Trump}      & 7                 & 7                 \\ \hline
\textbf{5G}         & 5                 & 4                 \\ \hline
\textbf{Bill Gates} & 7                 & 7                 \\ \bottomrule
\end{tabular}
\end{subtable}
\end{table}

Furthermore, we found that both non-credible bots and humans exhibit high commonality in retweeted domains (URLs) and retweeted users. 
Again, take the 5G topic as an example, the non-credible bots and humans shared many popular domains as well as users (Table~\ref{table:5G_common}).
The same analyses were also conducted for other topics. 
For each user type, Table \ref{tab:top15summary} summarizes the domains and users that commonly appeared in retweets in all the topics.
The non-credible bots shared many in common with respect to the top-10 retweeted domains and users, indicating the similarity in retweet behaviors between non-credible bots and humans (as well as credible bots and humans).
These findings further support the assumption that non-credible bots were following non-credible humans rather than credible accounts.

\section{Discussion}
In this paper, we investigated the roles of bots by analyzing retweet networks, temporal patterns of retweets as well as retweeted contents during the COVID-19 infodemic. 
For analysis, we focused on misinformation and conspiracy-theory-related topics, such as ``WHO'', ``Trump'', ``5G'', and ``Bill Gates''. 
Our analyses provided two major findings: the segregated retweet networks of bots and their temporal and topical similarities.
We revisit the findings and discuss their implications here.

First, we found that the retweet networks exhibited segregation and parasitic natures, suggesting two types of voices or echo chambers in all the topics (Fig.~\ref{fig:RT}).
One represents mainstream media and public institutions and the other represents right-wing (self-)media and celebrities. 
The echo chamber structures may amplify mis/disinformation from non-credible sources while hindering the dissemination of information from legitimate sources.  
According to the indegrees (Fig.~\ref{fig:botometer_vs_indegree}), the basic influence of retweets by non-credible humans can be much larger than that by non-credible bots. Thus, one might think that bots did not play as important a role during the COVID-19 infodemic as they did in previous political events, including the 2016 US presidential election. However, we cannot simply make this a definitive conclusion. 
Rather, the clustering of non-credible bots may reflect a partisan asymmetry and that non-credible bots follow non-credible humans. 
In particular, non-credible bots were parasitic on the far-right media and celebrity accounts, selectively amplifying their voices (Fig.~\ref{fig3}).
We speculate that non-credible bots targeted at such influential accounts because they needed to get amplification from prestigious accounts with social capital and high follower accounts and, thereby, extend their reach and engagement. The parasitic role of non-credible bots is for them to interact with accounts that might give them a chance to have their narratives or views amplified into the public sphere.

Second, our results show that retweet actions and contents were highly correlated between non-credible bots and humans for each topic (Figs.~\ref{fig:4topics} and \ref{fig:tfidf}). 
We found that non-credible bots preferentially distributed China-related terms (e.g., ``wuhan'',``china'', ``chinese'') from non-credible sites and users in the 5G topic (Fig.~\ref{fig:tfidf} and Table~\ref{table:5G_common}) and the other topics (Figs. 2a-f in SI). 
These consistent patterns may reflect that many (although not all) non-credible humans were diffusing China-related misleading claims and the non-credible bots were amplifying their effects. It is said that Trump’s fabrication of the term ``China virus'' led to racism targeted at Chinese and Asian Americans further fueled Asian hate and Sinophobia (anti-Chinese sentiment) in Western countries. Therefore, we suspect that bots were abused for spreading such propaganda during the early COVID-19 infodemic.
Although Twitter suspended many malicious accounts during the COVID-19 infodemic, sophisticated bots remained active and selectively parasitic on the partisan clusters.
Therefore, we conclude that the role of non-credible bots is non-negligible in the COVID-19 infodemic.

This evidence and these considerations cause us to rethink the weaponized use of bots for online manipulation and political interventions. 
This calls for the necessity of continuously monitoring the information ecosystem of bots. 
This is especially important to detect their coordinated acts, although we did not find evidence of such events in the current settings. However, this could still represent  a credible future threat that would likely have a negative societal impact. 
As the WHO mentioned, an infodemic is a ``second disease'' that emerged alongside with COVID-19, and it is important to take immediate action to address this infodemic.
As demonstrated in this study, social media analysis is important in order to acquire an overview of the infodemic and gain insights into a mitigation strategy. 

Our study has several limitations that need to be addressed in future research. Since Twitter suspends any accounts that it considers as ``malicious'', we were unable to obtain a comprehensive picture of users' interactive behaviors in this study. We also had limited information about the sources of credible and non-credible domains (URLs), which require frequent updates; thus, all of the URLs could not be labelled in our analyses. 
The availability problem of a credible/non-credible domain list requires a collective effort to solve.
Despite these limitations, this study furthers our understanding of the roles of bots in misinformation propagation during an infodemic in the midst of a world-wide healthcare crisis, and re-emphasizes the need to develop an efficient method to address behaviors of malicious bots. 


%
\section*{Conflict of interest}
The authors declare that they have no conflict of interest.



\bibliographystyle{spphys} 
\bibliography{refs}

\end{document}